%% file: main.tex
\documentclass[conference]{IEEEtran}

\usepackage{cite}
\usepackage{amsmath,amssymb,amsfonts}
\usepackage{algorithmic}
\usepackage{graphicx}
\usepackage{textcomp}
\usepackage{comment}
\usepackage[nolist]{acronym}
\usepackage[nolist]{acronym}
\usepackage[normalem]{ulem}
\usepackage{comment}
\usepackage{color,soul}
\usepackage{multirow}
\usepackage{multicol,multirow,colortbl}
\usepackage[caption=false,font=footnotesize,labelfont=sf,textfont=sf]{subfig}
\usepackage{algorithm}
\usepackage[dvipsnames]{xcolor}

\UseRawInputEncoding

\newcommand{\red}[1] {\textcolor[rgb]{1.0,0.0,0.0}{{#1}}}

\newcommand{\bazzi}[1]{\textcolor{blue}{#1}}

\begin{document}
\bstctlcite{IEEEexample:BSTcontrol}
%
\title{Prescriptive Manoeuvre Coordination for Intersections Without Stops} 
\title{Maneuver Coordination Using V2I to Improve Traffic Efficiency at Intersections}

\author{\IEEEauthorblockN{
Lorenzo Farina\IEEEauthorrefmark{1}\IEEEauthorrefmark{2},
Marco Rapelli\IEEEauthorrefmark{3}, 
Barbara Mav\'{i} Masini\IEEEauthorrefmark{4}\IEEEauthorrefmark{2}, 
Claudio Casetti\IEEEauthorrefmark{3},
Alessandro Bazzi\IEEEauthorrefmark{1}\IEEEauthorrefmark{2}
}\\
\IEEEauthorblockA{\IEEEauthorrefmark{1}DEI, Universit\`a di Bologna, 40136 Bologna, Italy}
\IEEEauthorblockA{\IEEEauthorrefmark{2}National Laboratory of Wireless Communications (WiLab), CNIT, 40136 Bologna, Italy}
\IEEEauthorblockA{\IEEEauthorrefmark{3}Department of Control and Computer Engineering (DAUIN), Politecnico di Torino, 10129 Torino, Italy}
\IEEEauthorblockA{\IEEEauthorrefmark{4} National Research Council (CNR) - IEIIT, 40136 Bologna, Italy}
}


\markboth{Journal of \LaTeX\ Class Files,~Vol.~14, No.~8, August~2015}%
{Shell \MakeLowercase{\textit{et al.}}: Bare Demo of IEEEtran.cls for IEEE Transactions on Magnetics Journals}
%




\input{Acronyms.tex}
\input{definitions.tex}

\IEEEtitleabstractindextext{%
\begin{abstract} 
Traffic efficiency and air pollutants are major issues in urban transport systems, as they are critical points for traffic flows and people living in urban areas, respectively.
In response to this challenge, connected and autonomous vehicles (CAVs) have garnered increasing attention, with the potential to address the above challenges through cooperative maneuvering and real-time communication. In this paper, we consider an intersection served by a multi-access edge computing (MEC) where CAVs communicate with a controller to pass the intersection without the need of signals and without stopping unless in presence of heavy congestion. For this purpose, we propose a first-in-first-scheduled (FIFS) algorithm, and we specify the communication protocol for its practical implementation. The solution is validated through simulations with varying traffic densities, showing that it provides significant gains in travel time and emissions compared to both uncontrolled intersections, with or without traffic lights, and a benchmark solution derived from the literature.\footnote{This is the accepted version of the paper. ©2024 IEEE. Personal use of this material is permitted. Permission from IEEE must be obtained for all other uses, in any current or future media, including reprinting/republishing this material for advertising or promotional purposes, creating new collective works, for resale or redistribution to servers or lists, or reuse of any copyrighted component of this work in other works.}
\end{abstract}
\begin{IEEEkeywords}
Maneuver coordination, intersection management, V2I, traffic efficiency.
\end{IEEEkeywords}
}

\maketitle

\IEEEdisplaynontitleabstractindextext

%
\IEEEpeerreviewmaketitle

\acresetall

\section{Introduction}

Lowering traffic congestion and pollutant emissions currently rank among the biggest challenges in urban areas. In addition to an estimated total of 54 extra hours per year spent by drivers in traffic congestion~\cite{schrank2019urban}, the World Health Organization (WHO)~\cite{world2018burden} reports 7 million deaths caused by air pollution every year.

Recently, \ac{CAV} technologies presented a promising avenue for improving traffic efficiency while mitigating pollutant emissions.
These solutions, leveraging advancements in artificial intelligence, sensor systems, and \ac{V2X} 
communications, offer the potential to revolutionize the management of intersections and other critical road segments by enhancing situational awareness, optimizing traffic flow, and reducing emissions.
Recognizing the urgency of addressing intersection issues, the research community has mobilized efforts at an international level to explore innovative solutions~\cite{MaksimovskiFestagFacchi2021,9678327,GUO2019313,10.1145/3407903,a15090326}. 

In this context, 
standardization entities such as the \ac{SAE} and the \ac{ETSI} are playing a pivotal role in shaping the landscape of \ac{V2X} standardization. 
In addition to established solutions for cooperative awareness and ongoing work on sensor sharing, an aspect that is recently receiving attention is the definition of standards to enable maneuver coordination. In particular, \ac{SAE} published the \ac{MSCS} in SAE J3186 and J3216~\cite{SAEJ3186,SAEJ3216}, while \ac{ETSI} is working at the \ac{MCS} in the ETSI TR 103 578 and TS 103 561~\cite{etsi103.578,etsi103.561}. These documents define the messages and protocols enabling vehicles to coordinate their actions and include the intersection as one of the main use cases. It is important to underline that these standards do not propose actual traffic management algorithms. 

On the one hand, standards are being defined to support coordinating messages; on the other hand, researchers are identifying algorithms to optimize the use of intersections with or without the presence of signals and traffic lights. Relevant studies are those about the optimization of traffic lights through the \ac{GLOSA} application~\cite{5982524} or about \acp{VTL}~\cite{6058657,JAME2022167,BAZZI201642}. More recently, looking at scenarios where only \acp{CAV} will be allowed to circulate in some areas, studies have been devoted to the case of signal-free intersections with maneuver coordination to pass the intersection~\cite{dresner2004multiagent,malikopoulos2018decentralized,zhang2019decentralized,zhang2018decentralized,wang2021digital,8264312}. 
However, most of the proposals in the literature either concentrate on the communication protocol or on the control algorithm to optimize the use of the intersection; in either case, the other aspect is simplified or not detailed.

In this work, we design and validate a cellular-based \ac{FIFS} algorithm that allows vehicles to cross an unregulated intersection without the need to stop, unless the intersection is heavily congested. 
The contribution of this work can be summarized as follows: 
\begin{itemize}
    \item we propose a novel \ac{FIFS} algorithm for scheduling vehicles at signal-free intersections which advances the existing literature by: \textit{(i)} also considering cars turning at the crossroads, which is not always considered in related work; \textit{(ii)} including a single trajectory definition schema for preventing vehicles to modify their trajectory multiple times before reaching the intersection; \textit{(iii)} allowing CAVs to independently choose their motion pattern for reaching the intersection, considering that every CAV has its own motion characteristics;
    \item we detail the communication protocol between the vehicles and an intersection controller to implement the proposed algorithm in practice;  \item we show the performance of the proposal through simulations in a realistic intersection with variable vehicle density, demonstrating reduced travel time and lower emissions.
\end{itemize}

\section{Related Works}\label{sec:SoA}

In recent years, several studies have started focusing on cooperative maneuvers at intersections, suggesting different crossing procedures.
Among relevant surveys on this topic,~\cite{MaksimovskiFestagFacchi2021} presents use cases for maneuver coordination in general, while~\cite{9678327,GUO2019313,10.1145/3407903} focus on cooperative intersection management, discussing the possible goals, the intersection modeling, the coordination architectures, the scheduling policies, the wireless technologies, as well as the challenges of supporting human-driven vehicles or security concerns and attack models.
Beyond surveys, a relevant group of technical works focused on signal-free intersections, trying to minimize the stopping of vehicles through their trajectory optimization. Among these, a reservation-based intersection control mechanism is proposed in~\cite{dresner2004multiagent}, in which the intersection is subdivided into slots that can be reserved upon request by vehicles with a \ac{FIFO} approach; the simulated vehicles, however, can only go straight at the intersection, and turns are not envisioned. The same simplification is applied in~\cite{malikopoulos2018decentralized}, which aims not only to minimize travel time but also to optimize energy in a decentralized way. A subsequent study~\cite{zhang2019decentralized} extends this framework, including also trajectories with turns.
Additional studies have then investigated the possibility to relax the FIFO structure and introduce re-sequencing: in~\cite{zhang2018decentralized} the authors foresee the computation of a new crossing sequence each time a new vehicle enters the control zone; in~\cite{wang2021digital}, a vehicle can reserve slots inside the intersection, and collision avoidance can be performed by connecting and cooperating with vehicles having conflicting paths. A comparison of different proposals in terms of travel time, energy consumption, computational time, and fairness was finally presented in~\cite{8264312}, where the considered strategies are differentiated in strategies with and without re-sequencing on the basis of their approach to the scheduling problem. As a conclusion, it was shown that re-sequencing is able to overcome some limitations of the FIFO approach; however, proposals that encompass the possibility of modifying the trajectory of a vehicle multiple times before it crosses the intersection could be a critical aspect when considering the design of communication protocol for a real implementation, which is an aspect not addressed in these works.

Unlike previous works, in this paper we design and validate a \ac{FIFS} algorithm and the related communication protocol, 
allowing vehicles to cross the intersection without the need to stop (unless the intersection is heavily congested), yet based on a single trajectory definition phase. 



\begin{figure}[!t]
\centering
\subfloat{\includegraphics[width=80mm]{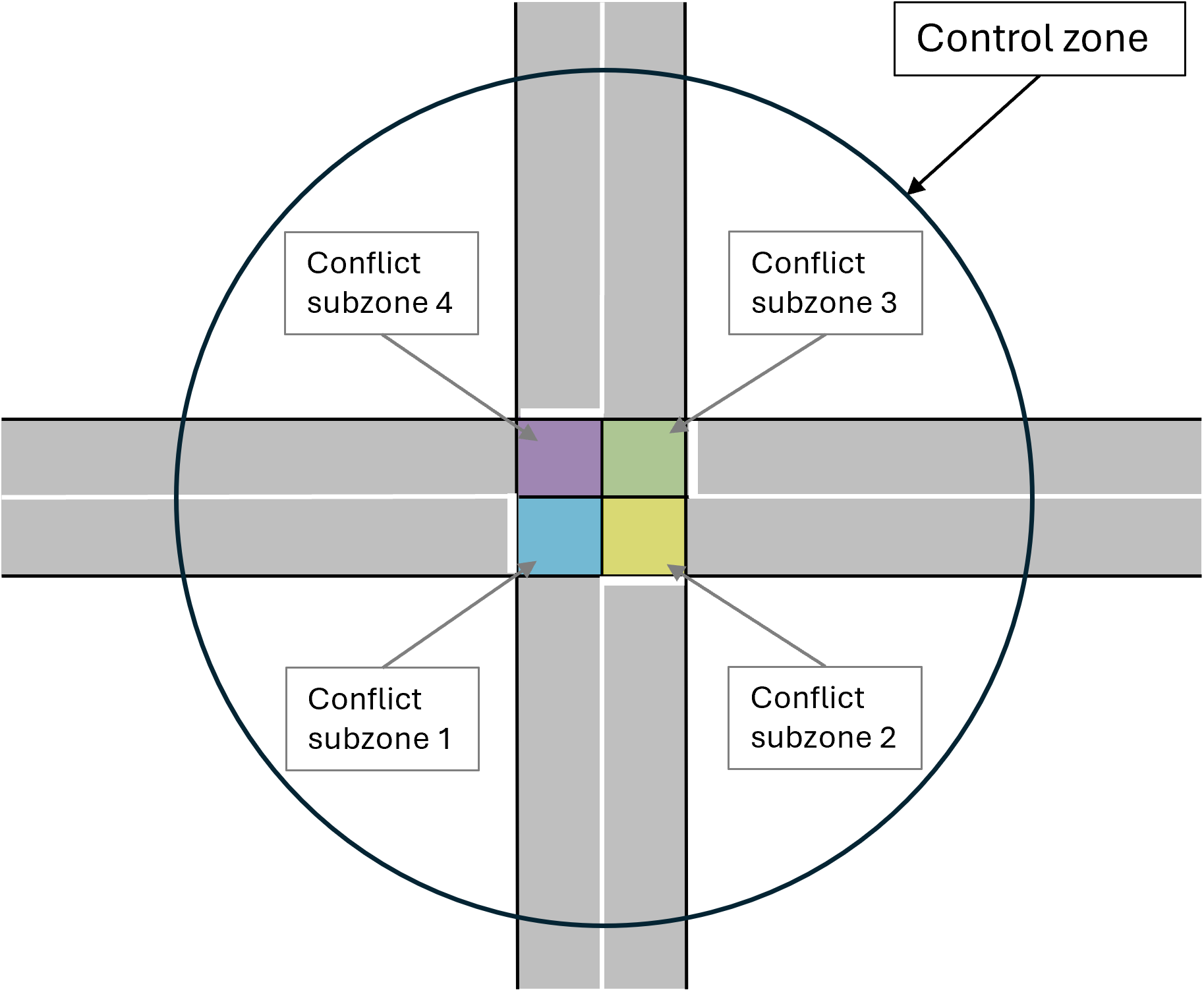}}
\caption{Intersection model with control zone and conflict subzones.}
\label{fig:model}
\end{figure}

\section{Model Design}\label{sec:Intersection_modelling}



Though we can claim that the proposed solution is applicable to different types of intersections, in this paper we will provide a detailed description of the algorithm focusing on a classic single-lane, four-way intersection. The extension of the algorithm applied to different intersection types is left for future investigation.

The intersection is organized in a control zone and in a conflict zone, which is in turn divided into conflict subzones, as depicted in Fig.~\ref{fig:model}.
The \textit{control zone} is the area within which the control algorithm applies; more specifically, when a vehicle enters the control zone the algorithm starts. The junction itself is then the so-called \textit{conflict zone}, which is the area where lateral collisions may occur. The conflict zone is further subdivided into \textit{conflict subzones}. Each subzone can be occupied by only one vehicle at a time, but one car occupies two subzones during the transition from one to another. 

Only CAVs operating with \ac{V2I} communication will be considered in our scenario.
The intersection area is assumed to be entirely covered by an infrastructure, for example, a \ac{RSU} or a cellular base station. 
The area is also served by a \ac{MEC} server, guaranteeing limited end-to-end delay and hosting the \textit{intersection controller}, or simply \textit{controller}. In particular, we envision having CAVs communicate with the controller using standard messages dedicated to coordinated maneuver, like the SAE or ETSI maneuver coordination messages~\cite{etsi103.578,etsi103.561,SAEJ3186} exploiting the concept of \ac{TRR} to reserve the conflict subzones as scheduled.


The trajectory definition is performed by each vehicle independently from other vehicles, following the scheduling imposed by the controller. This makes the algorithm easier to implement in practice since each CAV may have different characteristics and therefore operate in a different way.


Before detailing the algorithm in Section~\ref{sec:FIFS}, we introduce the control constraints applied in the model. In particular:
\begin{itemize}
    \item The initial position corresponds to the entrance of the control zone and is indicated as $\position{\instant{0}} = 0$~m, at time $\instant{0}$;
    \item The vehicle reaches the conflict zone at time $\tTAI$, with final position $\position{\tTAI} = \controlLength$; please note that $\controlLength$ is a fixed value for a given intersection indicating the control zone range, whereas $\tTAI$ depends on the output of the algorithm; 
    \item The speed is $\speed{t} \leq \speedmax$ $\forall t \in [\instant{0}, \tTAI]$, with $\speedmax$ being the road speed limit;
    \item When the car goes straight, $\speed{\tTAI}\leq \speedmax$; when the car turns right or left, $\speed{\tTAI}\leq \speedturn$, where $\speedturn<\speedmax$ is the maximum speed allowed when the vehicle turns to grant a safe and comfortable maneuver;
    \item The acceleration is $\accelerationmin \leq \acceleration{t} \leq \accelerationmax$ $\forall t \in [\instant{0}, \tTAI]$, with $\accelerationmin<0$ maximum deceleration due to braking and $\accelerationmax>0$ maximum acceleration, both depending on the capabilities of the specific vehicle.
\end{itemize}

\section{First-In First-Scheduled Algorithm}\label{sec:FIFS} 

The algorithm is initiated by each vehicle, hereafter \ac{EGO}, when it enters the control zone. Algorithm~\ref{algo1_test} describes the commands executed by the \ac{EGO}, while Algorithm~\ref{algo2_test} refers to the instructions performed by the controller. The proposed algorithm is based on the three parts detailed in the following subsections and a backup mode discussed at the end of the section.

\input{algorithm_ego}

\begin{figure*}[!t]
\subfloat[]{\includegraphics[width=0.65\columnwidth]{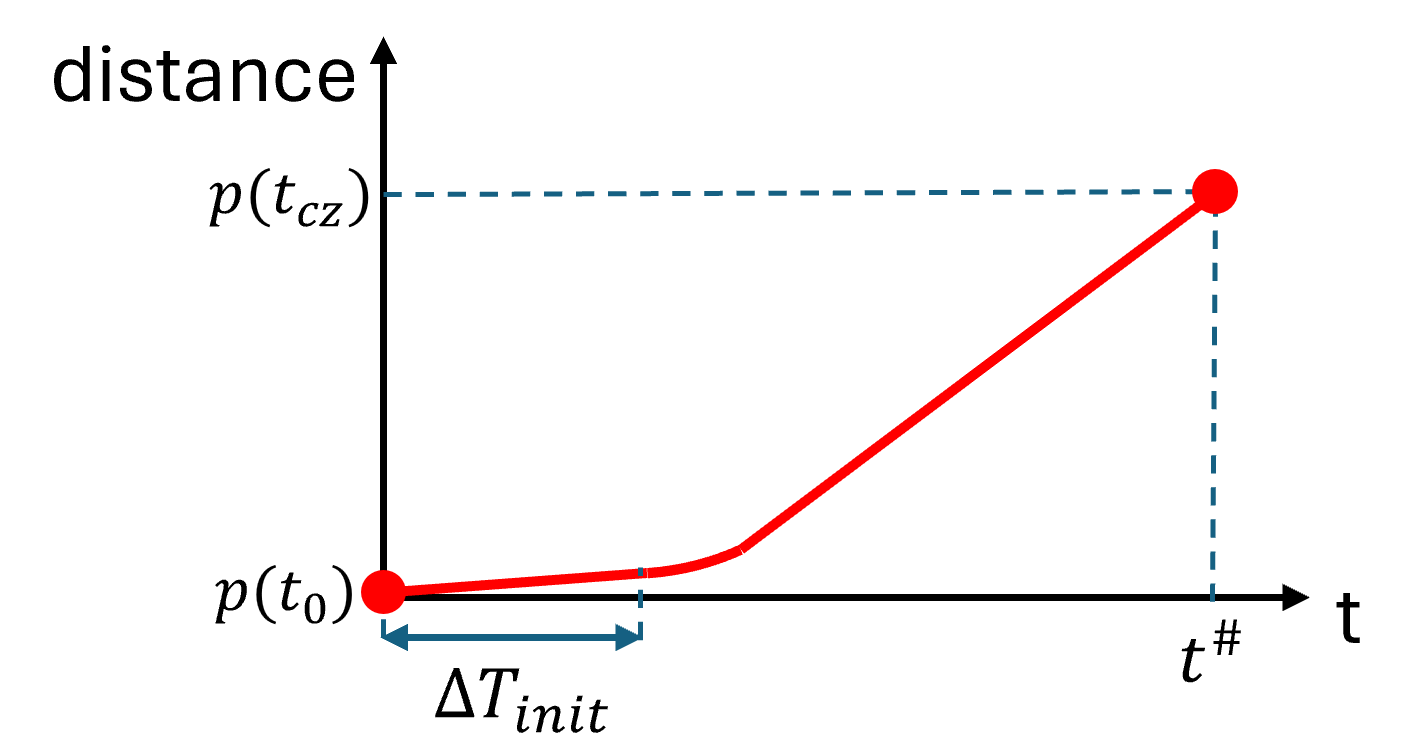}
\label{fig:ExampleTraj_pos}}
\hfill
\subfloat[]{\includegraphics[width=0.6\columnwidth]{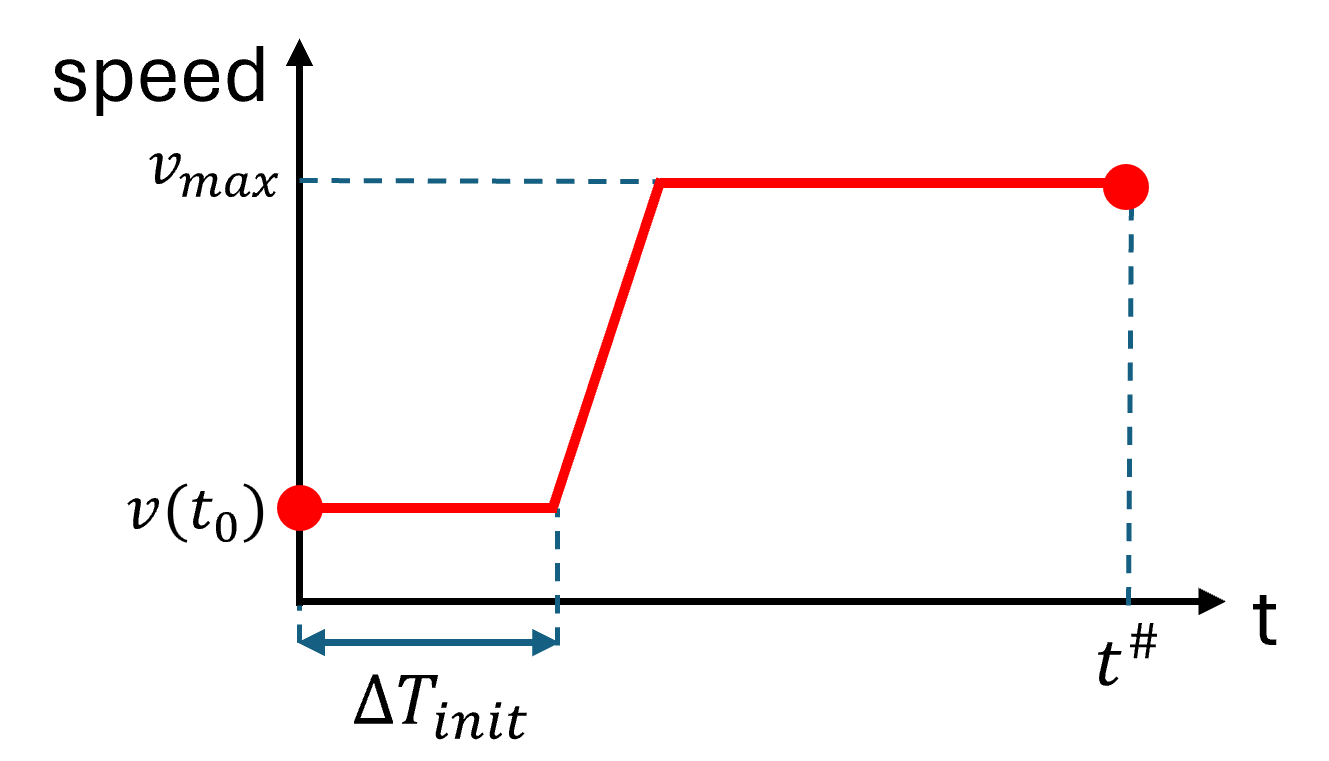}
\label{fig:ExampleTraj_speed}}
\hfill
\subfloat[]{\includegraphics[width=0.6\columnwidth]{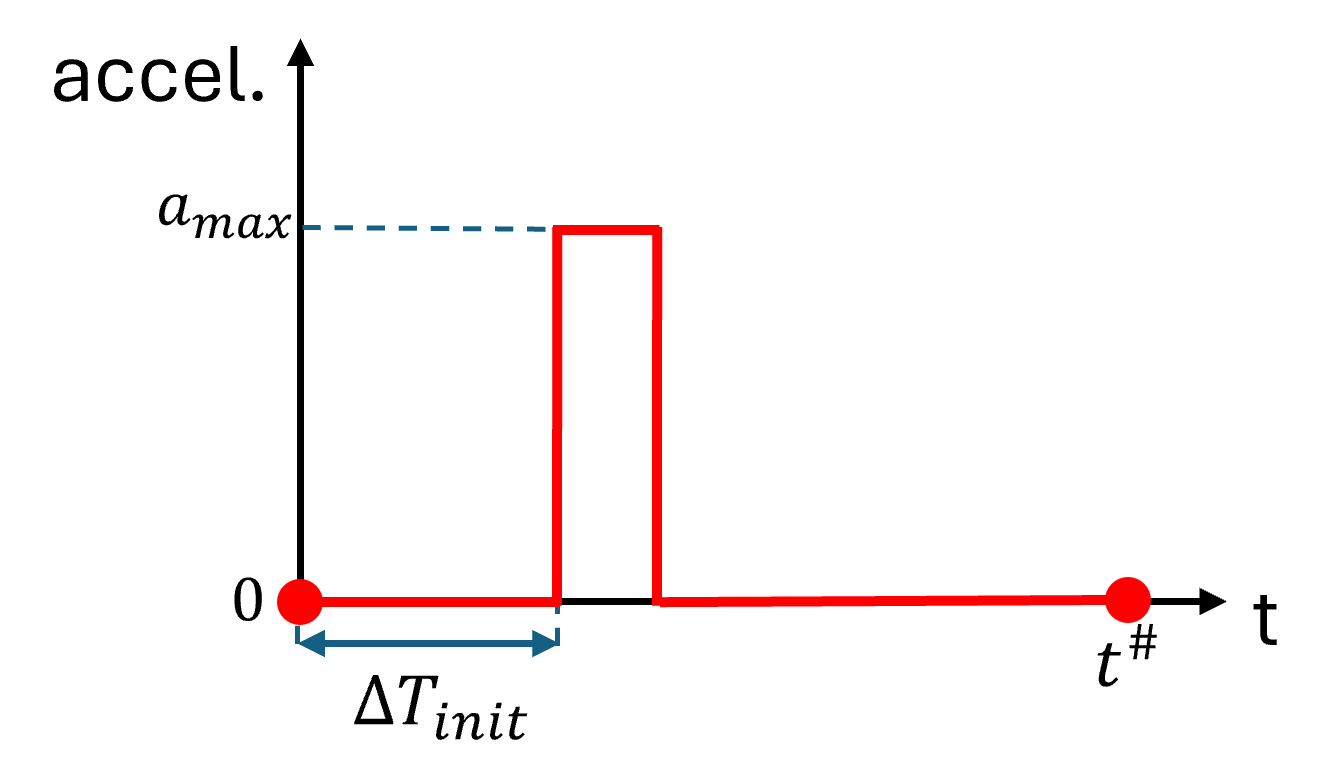}
\label{fig:ExampleTraj_acc}}
\caption{An example of trajectory planning performed by the vehicle with the setting of the motion profiles. (a) Example position profile. (b) Example speed profile. (c) Example acceleration profile.}
\label{fig:ExampleTraj}
\end{figure*}

\input{algorithm_controller}

\subsection{Part 1: Preliminary evaluation at the EGO}
In the first part, the \ac{EGO} calculates the minimum time to the intersection. Specifically, when entering the control zone, the EGO estimates the trajectory to reach the intersection, assuming that there are no other vehicles in the control zone (Alg.~\ref{algo1_test}, line 8). 
The calculation is performed assuming an initial time interval of duration $\deltaTinit$ during which the speed is kept constant; this interval has the objective of avoiding speed changes before the communication exchange between the EGO and the controller has concluded. The duration $\deltaTinit$ should be set according to the estimated delay introduced by the communication exchange. Once the trajectory is calculated, the corresponding time at which the conflict zone is reached is denoted as $\instant{\#}$.
In our implementation, the trajectory calculation minimizes the arrival time under the constraint to reduce both acceleration and deceleration. 
Indeed, it is a proven fact that driving at variable speeds leads to an increase in fuel consumption, hence an increase in pollutant emissions~\cite{TRL384}. 
Specifically, three cases are possible: \textit{(i)} if $\speed{\instant{0}}$ is lower than the target speed, $\instant{\#}$ is calculated assuming an initial acceleration (after $\deltaTinit$) up to the target speed and then maintained until the intersection; \textit{(ii)} if $\speed{\instant{0}}$ is the same as the target speed, $\instant{\#}$ is calculated assuming constant speed; \textit{(iii)} if $\speed{\instant{0}}$ is higher than the target speed, $\instant{\#}$ is calculated assuming that the deceleration starts at the latest possible instant, and the initial speed is maintained until that time. 
In Fig.~\ref{fig:ExampleTraj}, an example of the position, speed, and acceleration profile is shown in the case where the target speed is $\speedmax$ and $\speed{\instant{0}}<\speedmax$.

At this stage, the EGO calculates the minimum instant to enter the conflict zone (i.e., $\instant{\#}$) and the instants it would enter and exit from each subzone in such a case (the exiting subzone also indicates the direction after the conflict zone). The EGO sends now a proposal to the controller with all the calculated time instants; it also starts a timer with duration $\timeout$ and waits for the reply from the controller (Alg.~\ref{algo1_test}, lines 9-10). If the timer expires before a reply is received, the EGO recalculates the tentative trajectory from its current position, sends a new proposal, and restarts the timer (Alg.~\ref{algo1_test}, line 2 and line 11); this may occur because of too many errors in the communication or because the controller is serving other vehicles before the EGO. In case the distance from the EGO position to the conflict area falls below a minimum threshold during this phase, the EGO enters what we call \textit{backup mode}, detailed in Section~\ref{sec:backupmode} 
(Alg.~\ref{algo1_test}, lines 3-7).

\begin{figure*}[t]
\centering
\includegraphics[width=0.95\textwidth, draft=false]{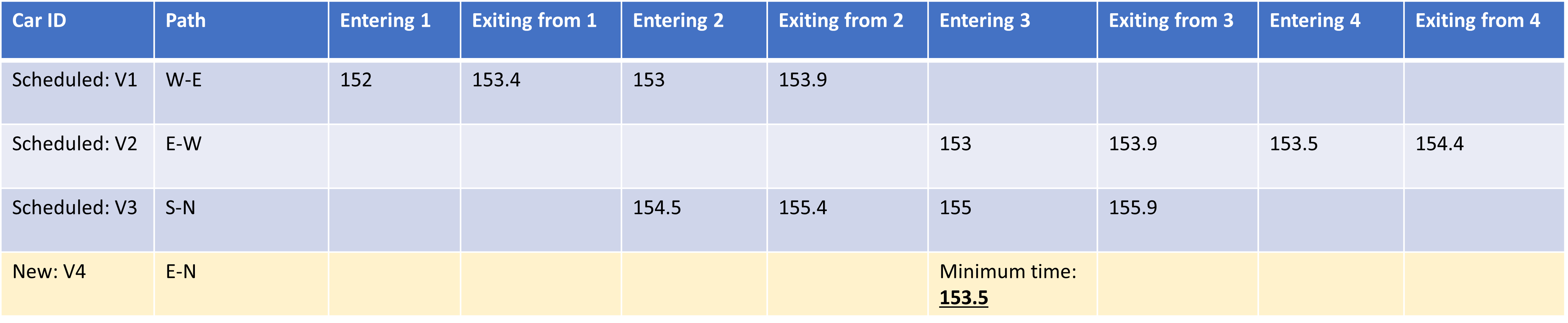}
\caption{Example of the scheduling table, where three vehicles (V1, V2 and V3) are already scheduled and the controller needs to schedule V4.}
\label{fig:table}
\end{figure*}


\begin{figure}[!b]
\centering
\vspace{-1.5\baselineskip}
\subfloat[]
{\includegraphics[width=0.8\columnwidth]{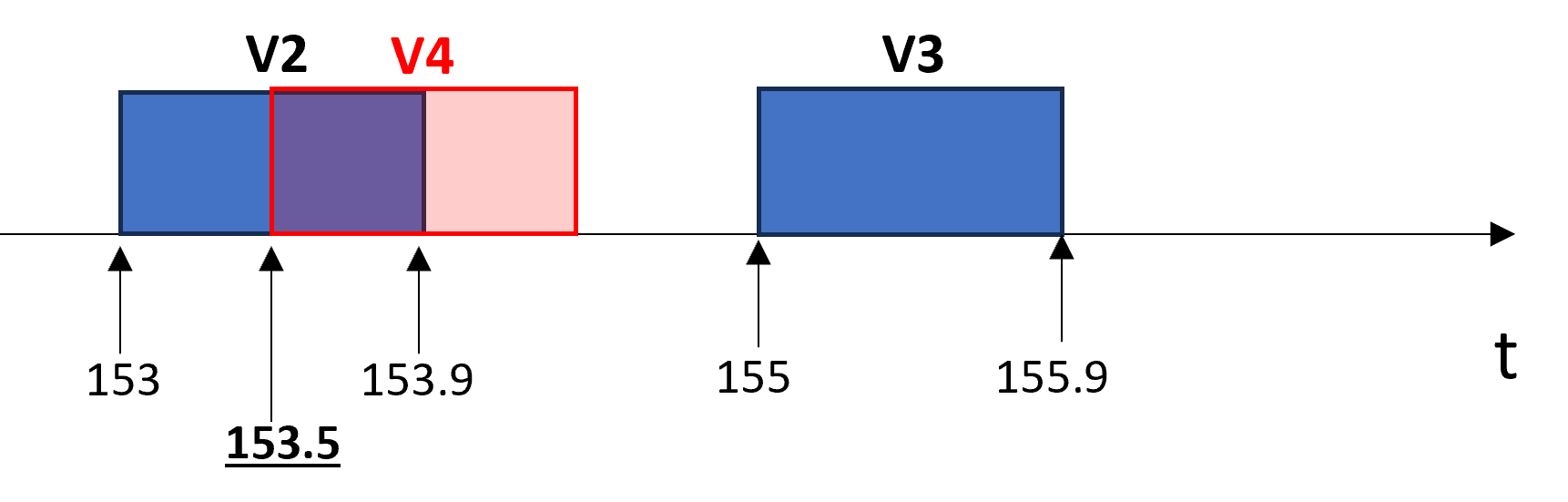}
\label{fig:Exampleconflict1}} \\
\subfloat[]{\includegraphics[width=0.8\columnwidth]{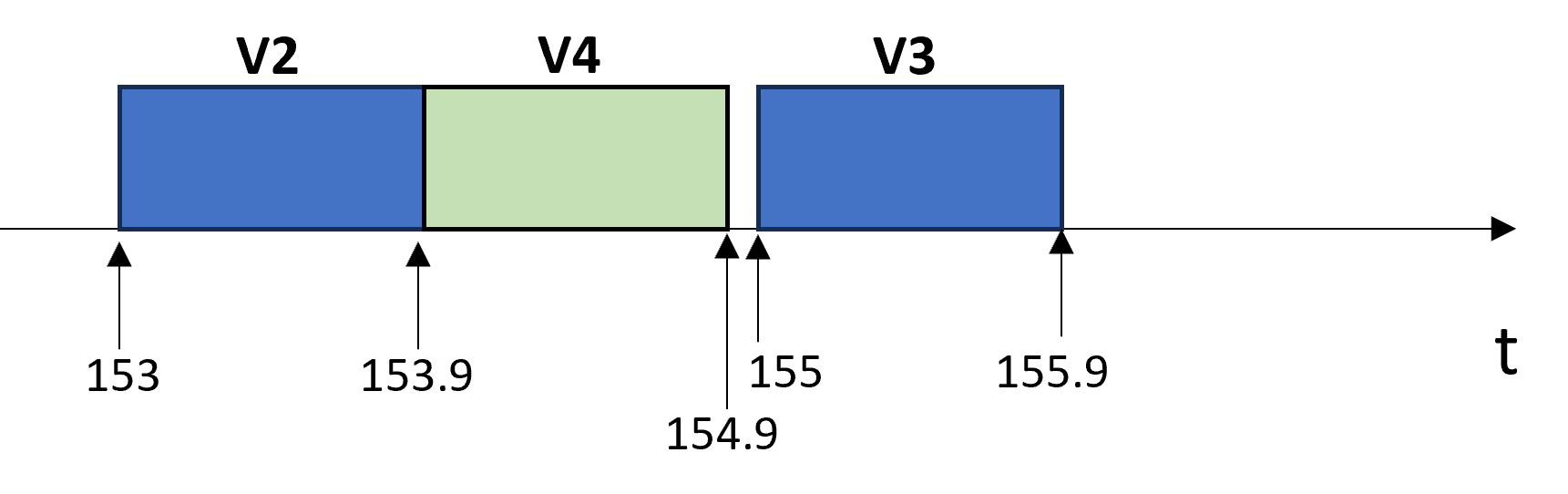}
\label{fig:Exampleconflict2}}
\caption{Example where a vehicle is scheduled between the other two. (a) Example where the minimum time of V4 conflicts with V2. (b) Example where V4 can be scheduled between V2 and V3.}
\label{fig:Exampleconflict}
\end{figure}

\subsection{Part 2: Evaluation and prescription from the controller} 

The controller verifies the proposal and, if needed, calculates different instants. This is done in three steps (Alg.~\ref{algo2_test}, lines 3-5).

\subsubsection{Part 2.A: The intersection controller checks for collisions before the conflict zone} 
The intersection controller needs to check that the motion profile of the EGO does not conflict with that of the vehicle ahead. In particular, it compares the estimated profiles of the two vehicles locally, and if the two vehicles are estimated to be closer than a minimum safety distance in any instant, the time to the intersection is increased.

In our implementation, this is done by comparing the estimated position of the two vehicles every 0.5~s 
and increasing the time to intersection by 0.2~s 
whenever the distance falls below 2~m. 
As a result, the longer the collision interval, the higher the time increase, leading to a safe position profile.

\subsubsection{Part 2.B: The intersection controller checks for collisions inside the intersection}
The aim of this step is to avoid side collisions inside the conflict zone by ensuring that there is always up to one vehicle in each subzone. 
To this end, once a vehicle terminates the algorithm and the instants it enters and exits from each subzone are fixed, these values are stored in a dynamic table, as exemplified in Fig.~\ref{fig:table}.  
Since vehicles are considered scheduled once the profiles are defined by the controller, the table can be updated only by removing one line when a vehicle exits the conflict zone and by adding a new line when a new vehicle is introduced. Notice that, since a vehicle can occupy more than a subzone at the same time, the entering instant of a subzone in the dynamic table always precedes the exit instant of the subsequent subzone.

Indeed, the EGO needs to go through one or more subzones sequentially, without interruptions. The intersection controller needs to find the first instant to enter the conflict zone, which allows the EGO to stay in each subzone when none of the scheduled vehicles occupy the same subzone. Calling the intervals of time when none of the scheduled vehicles occupy the subzone as \textit{idle intervals}, the intersection controller starts from the first subzone and searches for the first idle interval that allows the vehicle to cross the subzone; once found, it checks if this allows the vehicle to also cross the second and subsequent subzones during an idle interval; if this is not the case, the intersection controller goes back to the first subzone and searches for the next idle interval. 

As an example of this procedure, Fig.~\ref{fig:Exampleconflict} depicts a case consistent with subzone 3 in Fig.~\ref{fig:table}: in Fig.~\ref{fig:Exampleconflict1}, the initial time to intersection brings to a conflict between the EGO (V4) and a scheduled vehicle (V2); in Fig.~\ref{fig:Exampleconflict2}, the first idle interval is sufficient to avoid collisions with both the vehicles that are already scheduled through the same subsection (V2 and V3); if the idle interval between the reservations of V2 and V3 was not long enough, the EGO would have been scheduled after V3. In this latter case, if the EGO detects that the delay would not be enough to reach the intersection safely, it will send a contro-proposal to the intersection controller, or it will enter in \textit{backup mode}, detailed in Section~\ref{sec:backupmode}.



\subsubsection{Part 2.C: Intersection controller checks for collisions after the intersection} 
The purpose of the last step is to resolve conflicts arising at the exit of the intersection. This may happen when the EGO exits the conflict area at a speed higher than the vehicle ahead (i.e., the vehicle that, after the conflict area, precedes the EGO). 
In order to predict a possible collision after the intersection, the intersection controller computes \textit{(i)}~the distance traveled by the EGO vehicle, assuming that it uniformly slows down, before reaching the same speed of the vehicle ahead, denoted as $\dbrakeafter$; and \textit{(ii)}~the distance traveled by the vehicle ahead after the conflict zone, before the EGO reaches the distance $\dbrakeafter$, assuming that the vehicle ahead keeps a constant speed; this distance is denoted as $\dgapafter$; if $\dbrakeafter>\dgapafter-\dsafeafter$, where $\dsafeafter$ is a margin, then the time to intersection is increased. 
In such cases, in our implementation the time to intersection is incremented so that $\dbrakeafter=\dgapafter+\dsafeafter$, with $\dsafeafter=6$~m.


If any of the collision checks result in increasing the time to intersection for the EGO, the intersection controller repeats the whole calculation until no conflicts are found in the crossing path (Alg.~\ref{algo2_test}, line 2 and line 6).

Once all checks are successful, the intersection controller informs the EGO of the calculated instants of entrance and exit of each subzone (Alg.~\ref{algo2_test}, line 7). Please note that since the new time to intersection cannot be lower than $\instant{\#}$, the vehicle can surely reach the intersection at this new instant. 

\subsection{Part 3: Mobility profiles settings and crossing scheduling}

Once the EGO receives the prescription, it calculates the speed profile to satisfy the received time to intersection 
and confirms the plan by sending all the calculated time instants of entry and exit from the subzones (Alg.~\ref{algo1_test}, line 12 and lines 18-21). 
If the vehicle is unable to fulfill the prescription (meaning that it will surely reach the intersection later), 
it returns to the first part of the procedure and performs a new crossing proposal (Alg.~\ref{algo1_test}, line 1 and line 22). 

If the speed profile calculated by the EGO goes below a given threshold, or if an agreement is not reached before the EGO is closer to the conflict zone than a threshold, 
the EGO enters the backup mode detailed in Section~\ref{sec:backupmode} 
(Alg.~\ref{algo1_test}, lines 3-7 and lines 13-17). At the end of the communication procedure, the EGO not in backup mode has the motion profiles (i.e., position, speed, acceleration) defined until exiting from the conflict zone. 

The procedure ends when the controller receives a confirmation from the EGO; otherwise, it waits for a new proposal from the EGO (Alg.~\ref{algo2_test}, lines 8-13).

\subsection{Backup mode}\label{sec:backupmode}

If an agreement is not reached before the vehicle is at a minimum distance from the control zone to safely break and stop (which is set to $L/2$ in our implementation) or the calculated speed profile goes below a given threshold (set to 3~m/s in our implementation), the EGO enters the \textit{backup mode} (Alg.~\ref{algo1_test}, lines 3-7 and lines 13-17). In this case, it returns to normal operations, meaning that the CAV relies on its sensors and normal road rules without following the indications from the controller. 

Backup mode is a last resort, which occurs when the intersection is heavily congested. Once an EGO in backup mode crosses the control zone, it sends a notice to the controller so that, in the meantime, the controller is aware that there is a vehicle driving in backup mode. In this situation, the controller waits until all vehicles in backup mode cross the intersection before re-initializing the algorithm (Alg.~\ref{algo2_test}, lines 14-17).


\vskip 0.2cm

Please observe that the described protocol is designed so that: \textit{(i)}~the load in terms of wireless traffic is very limited; \textit{(ii)}~the decisions are taken sufficiently in advance compared to the center of the intersection, allowing to introduce backup mode procedures when needed; \textit{(iii)}~the communication always involves only one vehicle at a time, and once the vehicle is scheduled, it does not need to have additional interactions with the controller. Thus, the number of exchanges only linearly increases with the increase in the number of vehicles at the intersection.

\section{Simulation Results}\label{sec:simulation}


Results are obtained through simulations performed using the traffic simulator SUMO~\cite{SUMO2018} and a script written in Python which communicates with SUMO using the Traffic Control Interface (TraCI). The script continuously checks the position of all the vehicles and controls their behavior from the instant they enter the control zone to the instant they leave the intersection, unless vehicles enter backup mode. 
The main simulation settings are summarized in Table~\ref{Tab:Settings}.

\subsection{Simulation settings and outputs}

\subsubsection{Intersection definition}
\begin{figure}[!b]
\centering
\subfloat{\includegraphics[width=50mm]{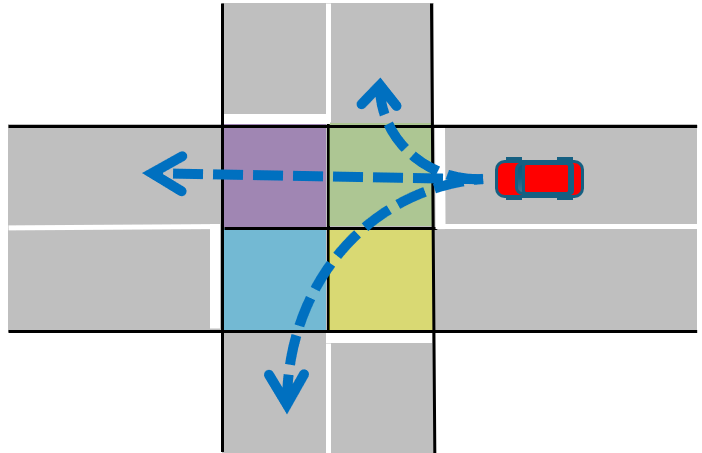}}
\caption{Intersection with subzones and possible EGO trajectories. }
\label{fig:exampledirections}
\vskip -0.3cm
\end{figure}

We consider an intersection with 4 converging roads, all assumed to be of length $2 \controlLength$, where $\controlLength$=100~m is the distance from the control zone limit to the beginning of the conflict zone. The conflict zone is a square with side 14.4~m, in turn composed of 4 equal subzones with side $7.2$~m. As represented in Fig.~\ref{fig:exampledirections}, the vehicle crosses one subzone when turning right, two subzones when going straight, and three subzones when turning left.
The vehicles are independently generated at the beginning of each road following a Poisson distribution with the same generation rate from each direction, and they turn left, turn right, or go straight with uniform probability.

\input{TableSettings.tex}

\subsubsection{Communication model} 
Regarding the modeling of communication, 
given the unicast nature and the assumption that there are no areas of outage around the intersection, it is reasonable to assume that all losses are detected and messages can be re-transmitted until they are correctly received. The communication impairments are therefore reproduced assuming a delay, which we model as uniformly randomly distributed between a minimum and a maximum value. In our simulation, the delay of each transmission from the EGO to the controller or from the controller to the EGO is modeled as a uniformly random variable between 20 and 100~ms, which are pessimistic values for 5G cellular systems with a \ac{MEC}.

\subsubsection{Benchmarks}
The performance of the proposed FIFS algorithm is assessed based on the following benchmarks:
\begin{itemize}
    \item \textit{Priority} case, where the horizontal roads have precedence over the vertical ones;
    \item \textit{Traffic light}, with a traffic light infrastructure with a green phase for streets facing each other of 35 seconds and a yellow phase of 3 seconds;
    \item \textit{FIFO} algorithm with ideal communication;
    \item \textit{FIFS} algorithm with ideal communication.
\end{itemize}
With \textit{ideal communications} we mean that the position of the vehicles is always perfectly known by the controller, and the decisions are instantaneously taken, without any delay.

\subsubsection{Output Metrics}
Performance is assessed through the following \acp{KPI}:

\begin{itemize}
    \item \textbf{Travel time}: time elapsed in seconds from the instant a vehicle enters the simulation to the instant it leaves;
    \item \textbf{Emissions}: CO$_2$ emissions from vehicles during their travel time, assuming the default settings of SUMO.
\end{itemize}

\subsection{Discussion}\label{sec:results}

\begin{figure}[!t]
\centering
\subfloat[]
{\includegraphics[width=0.9\columnwidth]{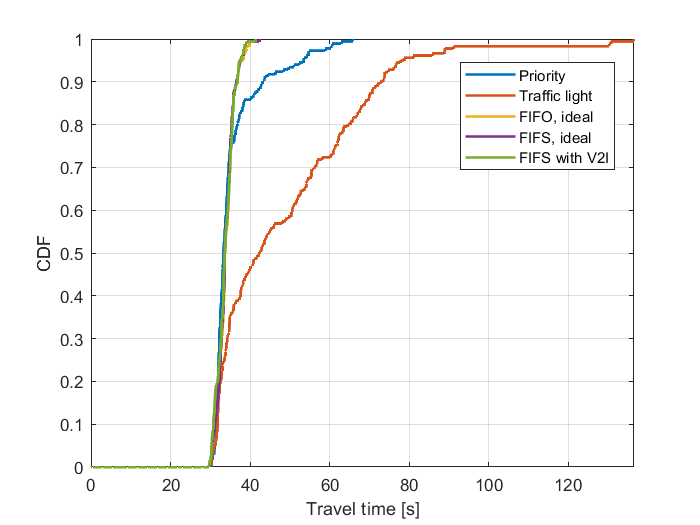}
\label{fig:Traveltime_0.05}} \\
\subfloat[]
{\includegraphics[width=0.9\columnwidth]{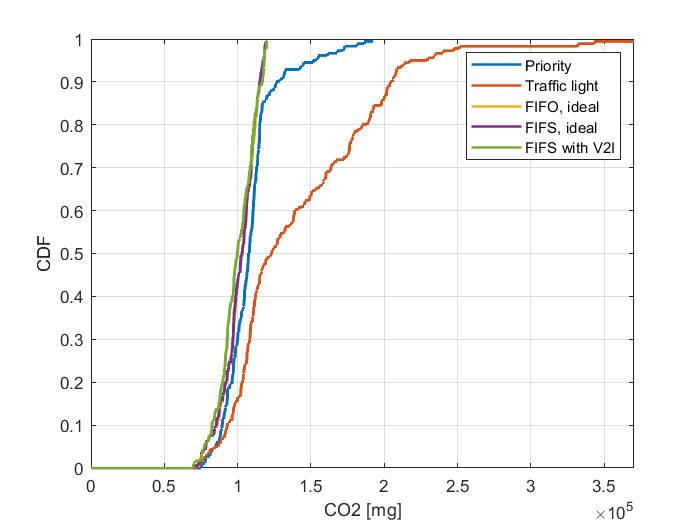}
\label{fig:Emission_0.05}}

\caption{Results with 0.05 vehicles/s per each direction. (a) Cdf of travel time. (b) Cdf of the emissions.}
\label{fig:Density0.05}
\end{figure}

\begin{figure}[!t]
\centering
\subfloat[]
{\includegraphics[width=0.9\columnwidth]{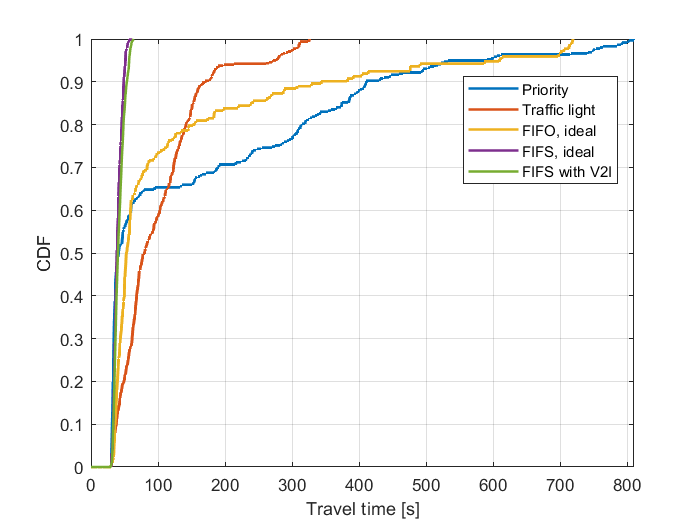}
\label{fig:Traveltime_0.15}} \\
\subfloat[]
{\includegraphics[width=0.9\columnwidth]{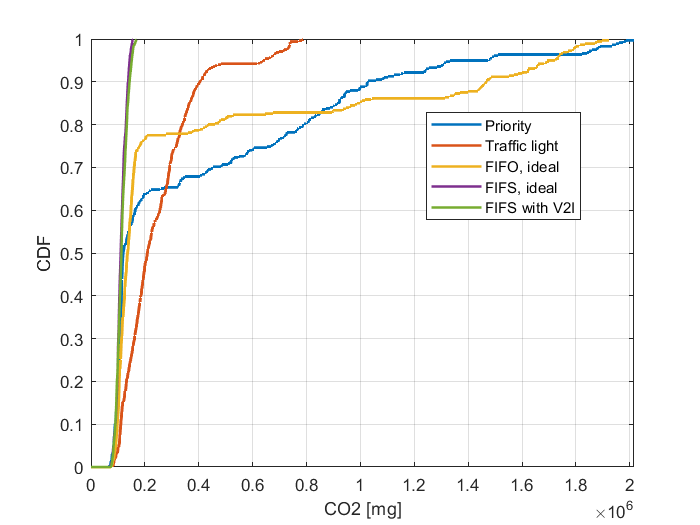}
\label{fig:Emission_0.15}}
\caption{Results with 0.15 vehicles/s per each direction. (a) Cdf of travel time. (b) Cdf of the emissions.}
\label{fig:Density0.15}
\end{figure}

\begin{figure}[!t]
\centering
\subfloat[]
{\includegraphics[width=0.9\columnwidth]{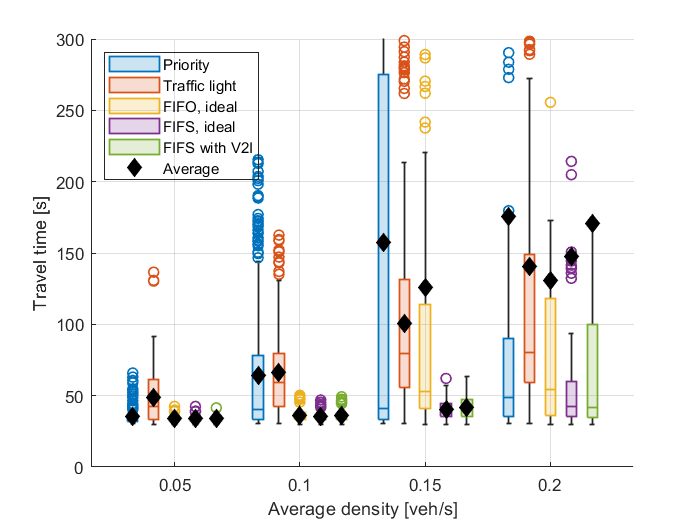}
\label{fig:boxchart_traveltime}} \\
\subfloat[]{\includegraphics[width=0.9\columnwidth]{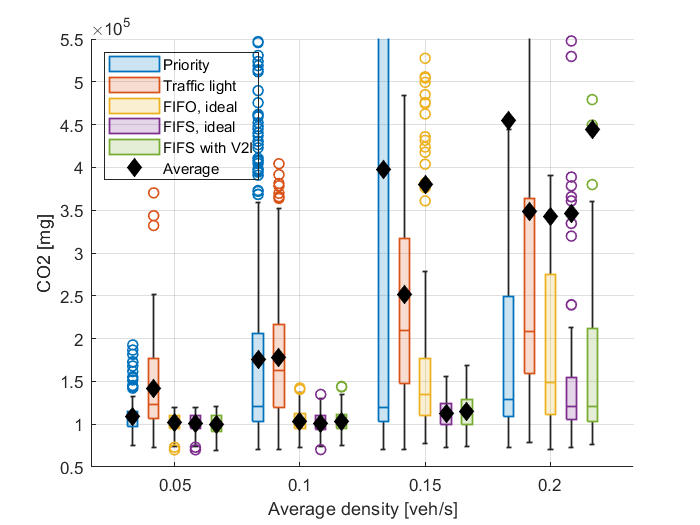}
\label{fig:boxchart_fuel}}
\caption{Box charts varying the density. (a) Travel time. (b) Emissions.}
\label{fig:boxchart}
\end{figure}

The outputs of the simulations are reported in Figs.~\ref{fig:Density0.05}-\ref{fig:boxchart}.
In Fig.~\ref{fig:Density0.05}, results refer to the case with an average input of 0.05 vehicles/s per direction, corresponding to a situation with relaxed traffic congestion. In particular, the \ac{cdf} of the travel time is shown in Fig.~\ref{fig:Traveltime_0.05} and the \ac{cdf} of the emissions is shown in Fig.~\ref{fig:Emission_0.05}. As is observable, when few vehicles are on the road, the use of the traffic light corresponds to the worst case, since it causes unnecessary stops and waiting times, which affect both output metrics. The travel time and emissions are significantly lower when one road has priority over the other; when looking at the travel time (Fig.~\ref{fig:Traveltime_0.05}), it can be observed that approximately 20\% of vehicles suffer from some delay compared to the others because they arrive from the road giving the right of way while the main road is not empty. All solutions with controlled \acp{CAV}, i.e., those implementing the FIFO and FIFS approaches, avoid stops at the intersection and yield the best performance. Under these conditions, results with FIFS and FIFO are very similar, and the communication protocol is shown to negligibly affect the performance.

In Fig.~\ref{fig:Density0.15}, the same plots are provided assuming an average input of 0.15 vehicles/s per direction, which corresponds to a situation with moderate traffic conditions. In this case, the traffic light solution is shown to overall outperform the case with priorities, since it avoids that a portion of the vehicles remains stopped at the intersection for long intervals. The FIFO approach shows instead its limits; even if the communications and decisions are ideal, the assumption that vehicles are allowed to enter the subzones only after the preceding vehicles are scheduled causes high travel times. Conversely, the FIFS algorithm allows all CAVs to be scheduled in an efficient way, with the travel time per vehicle that never exceeds 50~s and the CO$_2$ per vehicle that never exceeds 200~g. This holds also when \ac{V2I} realistic delays are considered.

Finally, Fig.~\ref{fig:boxchart} shows the box charts of the travel time and emissions varying the vehicle density. The box charts highlight  the upper and lower quartiles with a colored rectangle, while the vertical black lines highlight the range of the values except the outliers, indicated by circles. Black diamonds indicate the average values. 
The density in Fig.~\ref{fig:boxchart} varies from a minimum of 0.05 vehicles/s per direction (corresponding to the results shown in Fig.~\ref{fig:Density0.05}) to a maximum of 0.20 vehicles/s per direction. The latter case is a situation of high congestion, where queues are generated at the intersection. Similarly to what has already been observed and known from traffic theory, the solution with priorities is preferable to the one with the traffic light as long as the traffic congestion is moderate, and the reverse is true for congested traffic situations (at 0.20 vehicles/s, adopting priorities, some vehicles wait for very long time and thus the average travel time and emissions increase significantly). The FIFS algorithm with ideal communication 
shows always optimum performance at any density, both looking at the average results and worst cases. 
Even the FIFS algorithm with the designed communication protocol and realistic V2I delay, significantly outperforms the benchmarks without CAVs (i.e., priority and traffic light) and the FIFO scenario. Only when 0.20 vehicles/s per direction are introduced in the scenario, in the proposed solution a large number of vehicles enter in the backup mode, resulting in non-optimal performance. 



\section{Conclusion}\label{sec:conclusions}
In this paper, we addressed traffic efficiency and emission reduction for \acp{CAV} at unregulated intersections. Specifically, we proposed a \ac{FIFS} algorithm and a communication protocol that allow vehicles to safely cross an intersection served by a MEC without stopping thanks to the \ac{V2I} communication between vehicles and a controller. Results demonstrate the advantages of the proposed solution in terms of travel time and emissions with respect to intersections managed with priority, traffic lights, and FIFO algorithms. 

The proposed algorithm is general enough to be applied to intersections with different numbers of converging roads or lanes and to roundabouts. This aspect, together with the investigation of the applicability of human-driven vehicles, will be addressed in future works. Similarly, we do not consider here the presence of other non-connected vehicles and \acp{VRU}. In any case, a simple first approach for non-connected objects and \acp{VRU} could be the use of cameras to identify and notify them to the \acp{CAV} to fall back to normal operations.
%
%
%
%
Finally, another aspect that will deserve future activities is the adaptation to a distributed scenario with V2V communication rather than a centralized one with only V2I. In this way, we also envision to test the presented algorithm through a field test campaign to assess the performance in a real-world scenario.

\section*{Acknowledgment} This work was partially supported by the European Union under the Italian National Recovery and Resilience Plan (NRRP) of NextGenerationEU, partnership on “Telecommunications of the Future” (PE00000001 - program “RESTART”), project MoVeOver and was partially funded by the European Union through the project CONNECT under grant agreement no. 101069688.

\bibliographystyle{IEEEtran}  
\bibliography{biblio}

\end{document}

%% file: acronyms.tex
\begin{acronym} 
\acro{5G}{Fifth Generation of cellular networks}
\acro{5G NR}{5G new radio}
\acro{6G-V2X}{6G-based vehicle-to-everything}
\acro{BSS}{basic service set}
\acro{C-ITS}{cooperative-intelligent transport systems}
\acro{C-V2X}{cellular-\ac{V2X}}
\acro{CAM}{cooperative awareness message}
\acro{CAV}{connected and autonomous vehicle}
\acro{cdf}{cumulative distribution function}
\acro{CP}{cyclic prefix}
\acro{CPM}{collective perception message}
\acro{CPS}{collective perception service}
\acro{CSMA/CA}{carrier sense multiple access with collision avoidance}
\acro{D2D}{device-to-device}
\acro{DCM}{dual carrier modulation}
\acro{DSRC}{dedicated short range communication}
\acro{EED}{end-to-end delay}
\acro{EGO}{ego-vehicle}
\acro{ETSI}{European Telecommunications Standards Institute}
\acro{FCC}{federal communications commission}
\acro{FIFO}{first-in-first-out}
\acro{FIFS}{first-in-first-scheduled}
\acro{GLOSA}{green light optimised speed advisory}
\acro{GNSS}{global navigation satellite system}
\acro{KPI}{key performance indicator} 
\acro{LDPC}{low-density parity-check}
\acro{LOS}{line-of-sight}
\acro{LTE}{long term evolution}  
\acro{LTE-V2X}{LTE-based vehicle-to-everything} 
\acro{MAC}{medium access control}
\acro{MEC}{multi-access edge computing}
\acro{MCM}{manoeuvre coordination message}
\acro{MC}{manoeuvre coordination}
\acro{MCS}{manoeuvre coordination service}
\acro{MSCS}{maneuver sharing and coordinating service}
\acro{NLOS}{non-line-of-sight}
\acro{OBU}{on-board unit}
\acro{OCB}{outside of the context of a basic service set}
\acro{OFDM}{orthogonal frequency-division multiplexing}
\acro{PHY}{physical}
\acro{PoC}{proof-of-concept}
\acro{PRB}{physical resource block}
\acro{QAM}{quadrature amplitude modulation}
\acro{QoS}{quality of service}
\acro{RB}{resource block}
\acro{RRI}{resource reservation interval}
\acro{RSU}{road-side unit} 
\acro{SAE}{Society of Automotive Engineering}
\acro{SC-FDMA}{single carrier frequency division multiple access}
\acro{SCS}{subcarrier spacing}
\acro{SCI}{sidelink control information}
\acro{SoA}{state-of-the-art}
\acro{SoW}{statement of the work}
\acro{SPS}{semi-persistent scheduling}
\acro{TAI}{time at intersection}
\acro{TRR}{target road resource}
\acro{UE}{user equipment}
\acro{V2I}{vehicle-to-infrastructure} 
\acro{V2V}{vehicle-to-vehicle} 
\acro{V2X}{vehicle-to-everything} 
\acro{VAM}{vulnerable road user awareness message}
\acro{VRU}{vulnerable road user}
\acro{VTL}{virtual traffic light}
\acro{WAVE}{wireless access in vehicular environment}
\end{acronym}

%% file: definitions.tex
\newcommand{\controlLength}{L}
\newcommand{\position}[1]{p({{#1}})}
\newcommand{\speed}[1]{v({{#1}})}
\newcommand{\speedmax}{v_{\text{max}}}
\newcommand{\speedturn}{v_{\text{max-turn}}}
\newcommand{\speedAHEAD}[1]{v_\text{AHEAD}({{#1}})}
\newcommand{\speedEGO}[1]{v_\text{EGO}({{#1}})}
\newcommand{\acceleration}[1]{u({{#1}})}
\newcommand{\accelerationmin}{b_{\text{max}}}
\newcommand{\accelerationmax}{a_{\text{max}}}
\newcommand{\accelerationOne}{u_{\text{1}}}
\newcommand{\accelerationTwo}{u_{\text{2}}}
\newcommand{\accelerationThree}{u_{\text{3}}}
\newcommand{\instant}[1]{t_{{#1}}}
\newcommand{\tTAI}{t_\text{CZ}}
\newcommand{\deltaTinit}{\Delta T_\text{init}}
\newcommand{\dsafe}{d_\text{safe-before}}
\newcommand{\TaddAhead}{\tau_\text{add-before}}
\newcommand{\dgapafter}{d_\text{after}}
\newcommand{\dbrakeafter}{d_\text{brake}}
\newcommand{\tbrakeafter}{t_\text{brake}}
\newcommand{\dsafeafter}{d_\text{safe-after}}
\newcommand{\timeout}{\tau_\text{out}}
\newcommand{\braking}{b}
\newcommand{\delaymin}{D_\text{min}}
\newcommand{\delaymax}{D_\text{max}}

%% file: algorithm_EGO.tex
\begin{algorithm}[!b]
 \caption{EGO part of FIFS}
 \begin{algorithmic}[1]
   \renewcommand{\algorithmicrequire}{\textbf{Executed by the EGO}}
 \REQUIRE  when entering the control zone \\

  \REPEAT
    \REPEAT
      \IF{distance below safety threshold}
        \STATE Enter backup mode
        \STATE Send notice to the controller after the conflict zone
        \STATE EXIT(fail)
      \ENDIF
      \STATE Calculate the minimum time to the intersection based on current position
      \STATE Send proposal to the controller
      \STATE Start a timer $\timeout$ and wait for response
    \UNTIL{prescription not received from the controller}
    \IF {the prescription can be adhered to}   
      \IF {the speed profile falls below a threshold}
        \STATE Enter backup mode
        \STATE Send notice to the controller after the conflict zone
        \STATE EXIT(fail)
      \ENDIF
      \STATE Send confirmation to the controller
      \STATE Set the trajectory
      \STATE EXIT(success)
    \ENDIF
  \UNTIL{proposed prescription  can be adhered to}
 \end{algorithmic}\label{algo1_test}
\end{algorithm}

%% file: algorithm_controller.tex
\begin{algorithm}[!b]
 \caption{Controller part of FIFS}
 \begin{algorithmic}[1]
   \renewcommand{\algorithmicrequire}{\textbf{Executed by the intersection controller}}
  \REQUIRE  when a proposal is received \\
    \REPEAT
      \REPEAT
        \IF {conflicts identified before, inside or after the conflict zones}
          \STATE Postpone to the first acceptable time at intersection
        \ENDIF
      \UNTIL{no conflicts in the crossing proposal}
      \STATE Send prescription to the EGO
      \IF {confirmation is received from the EGO}
        \STATE Update the scheduling table
        \STATE EXIT(success)
      \ELSE      
        \STATE Wait for a new proposal from the EGO
      \ENDIF
      \IF {the EGO notifies a backup mode}
        \STATE Waiting for the EGO notice after the conflict zone
        \STATE EXIT(fail)
      \ENDIF   
  \UNTIL{confirmation is received from the EGO}   
 \end{algorithmic}\label{algo2_test}
\end{algorithm}

%% file: TableSettings.tex
\begin{table}[
t]
\caption{Main simulation parameters and settings. 
}\label{Tab:Settings}
\footnotesize
\centering
\begin{tabular}{p{4.5cm}p{3.5cm}}
\hline \hline
\textbf{Parameter} & \textbf{Value} \\  
Duration of each simulation & 1000 s\\
Road layout & 4 roads, one lane each \\
Lane width & 3.2 m \\
Control zone range $\controlLength$ & 100 m\\
Road length $2 \controlLength$ & 200 m\\
Conflict zone side & 14.4 m\\
Conflict subzone side & 7.2 m\\
Vehicle arrival distribution & Poisson, variable average\\
Maximum speed $\speedmax$ & 50 km/h \\ 
Maximum turn speed $\speedturn$ & 20 km/h \\ 
Maximum acceleration $\accelerationmax$ & 2.6 m/s$^2$\\
Maximum braking $\accelerationmin$ & -4.5 m/s$^2$\\
Initial interval $\deltaTinit$ & 500~ms \\
Communication delay (per message) & Between 20 and 100~ms\\
Safe margin after the intersec. $\dsafeafter$ & 6 m\\
Timer for the crossing proposal $\timeout$ & 500~ms \\
\hline \hline
\end{tabular}
\end{table}

%% file: biblio.bib
@IEEEtranBSTCTL{IEEEexample:BSTcontrol,
  CTLdash_repeated_names = "no"
}

@misc{schrank2019urban,
  title={Urban Mobility Report (Texas A\&M Transportation Institute, 2019)},
  author={Schrank, David and Lomax, T and Eisele, B},
  year={2019}
}

@techreport{world2018burden,
    author={World Health Organization WHO},
    title={Burden of disease from household air pollution for 2016},
    institution={World Health Organization WHO},
    year={2018} 
}

@inproceedings{MaksimovskiFestagFacchi2021,
  author    = {Maksimovski, Daniel and Festag, Andreas and Facchi, Christian},
  title     = {A Survey on Decentralized Cooperative Maneuver Coordination for Connected and Automated Vehicles},
  booktitle   = {Proceedings of the 7th International Conference on Vehicle Technology and Intelligent Transport Systems},
  publisher = {Scitepress},
  address   = {Set{\'u}bal},
  isbn      = {978-989-758-513-5},
  issn      = {2184-495X},
  doi    = {https://doi.org/10.5220/0010442501000111},
  pages     = {100 -- 111},
  year      = {2021},
}

@ARTICLE{9678327,
  author={Gholamhosseinian, Ashkan and Seitz, Jochen},
  journal={IEEE Access}, 
  title={A Comprehensive Survey on Cooperative Intersection Management for Heterogeneous Connected Vehicles}, 
  year={2022},
  volume={10},
  number={},
  pages={7937-7972},
  doi={10.1109/ACCESS.2022.3142450}}

@article{GUO2019313,
title = {Urban traffic signal control with connected and automated vehicles: A survey},
journal = {Transportation Research Part C: Emerging Technologies},
volume = {101},
pages = {313-334},
year = {2019},
issn = {0968-090X},
doi = {https://doi.org/10.1016/j.trc.2019.01.026},
url = {https://www.sciencedirect.com/science/article/pii/S0968090X18311641},
author = {Qiangqiang Guo and Li Li and Xuegang {(Jeff) Ban}},
}

@article{10.1145/3407903,
  title={A survey on intersection management of connected autonomous vehicles},
  author={Khayatian, Mohammad and Mehrabian, Mohammadreza and Andert, Edward and Dedinsky, Rachel and Choudhary, Sarthake and Lou, Yingyan and Shirvastava, Aviral},
  journal={ACM Transactions on Cyber-Physical Systems},
  volume={4},
  number={4},
  pages={1--27},
  year={2020},
  publisher={ACM New York, NY, USA}
}

@article{a15090326,
  title={Autonomous Intersection Management by Using Reinforcement Learning},
  author={Karthikeyan, P. and Chen, Wei-Lun and Hsiung, Pao-Ann},
  journal={Algorithms},
  volume={15},
  number={9},
  year={2022},
  doi={10.3390/a15090326}
}

@Article{electronics10141702,
AUTHOR = {Chavoshi, Kimia and Genser, Alexander and Kouvelas, Anastasios},
TITLE = {A Pairing Algorithm for Conflict-Free Crossings of Automated Vehicles at Lightless Intersections},
JOURNAL = {Electronics},
VOLUME = {10},
YEAR = {2021},
NUMBER = {14},
ARTICLE-NUMBER = {1702},
URL = {https://www.mdpi.com/2079-9292/10/14/1702},
ISSN = {2079-9292},
}

@ARTICLE{9800916,
  author={Chen, Chaoyi and Xu, Qing and Cai, Mengchi and Wang, Jiawei and Wang, Jianqiang and Li, Keqiang},
  journal={IEEE Transactions on Intelligent Transportation Systems}, 
  title={Conflict-Free Cooperation Method for Connected and Automated Vehicles at Unsignalized Intersections: Graph-Based Modeling and Optimality Analysis}, 
  year={2022},
  volume={23},
  number={11},
  pages={21897-21914},
  doi={10.1109/TITS.2022.3182403}}

@ARTICLE{8370718,
  author={Fayazi, Seyed Alireza and Vahidi, Ardalan},
  journal={IEEE Transactions on Intelligent Vehicles}, 
  title={Mixed-Integer Linear Programming for Optimal Scheduling of Autonomous Vehicle Intersection Crossing}, 
  year={2018},
  volume={3},
  number={3},
  pages={287-299},
  doi={10.1109/TIV.2018.2843163}}

@ARTICLE{9735394,
  author={Hadjigeorgiou, Andreas and Timotheou, Stelios},
  journal={IEEE Transactions on Intelligent Vehicles}, 
  title={Real-Time Optimization of Fuel-Consumption and Travel-Time of CAVs for Cooperative Intersection Crossing}, 
  year={2023},
  volume={8},
  number={1},
  pages={313-329},
  doi={10.1109/TIV.2022.3158887}}

@misc{pan2023hierarchical,
      title={A Hierarchical Robust Control Strategy for Decentralized Signal-Free Intersection Management}, 
      author={Xiao Pan and Boli Chen and Li Dai and Stelios Timotheou and Simos A. Evangelou},
      year={2023},
      eprint={2206.14986},
      note={arXiv 2206.14986},
      primaryClass={math.OC}
}

@INPROCEEDINGS{8264312,
  author={Zhang, Yue and Malikopoulos, Andreas A. and Cassandras, Christos G.},
  booktitle={2017 IEEE 56th Annual Conference on Decision and Control (CDC)}, 
  title={Decentralized optimal control for connected automated vehicles at intersections including left and right turns}, 
  year={2017},
  volume={},
  number={},
  pages={4428-4433},
  doi={10.1109/CDC.2017.8264312}}

@misc{TRL384,
	title={{TRL REPORT 384:} Reducing
the Environmental Impact of Driving: a Review of Training and in-Vehicle
Technologies},
	year={1999},
author={J. Cloke and G. Harris and S. Latham and A. Quimby and L. Smith and C. Baughan}
	}

@ARTICLE{etsi103.578,
author={ETSI}, 
journal={{ETSI} {TR} 103 578},
title={{Intelligent Transport Systems (ITS); Vehicular Communications; Informative report for the Maneuver Coordination Service}},
year={2023},
}

@ARTICLE{etsi103.561,
author={ETSI}, 
journal={{ETSI} {TS} 103 561},
title={{TS 103 561: “Intelligent Transport Systems (ITS); Vehicular Communications; Basic Set of Applications; Maneuver Coordination Service”}},
year={2023},
}

@ARTICLE{SAEJ3186,
author={SAE}, 
journal={{SAE} {J3216} },
title={Application Protocol and Requirements for Maneuver Sharing and Coordinating Service},
year={2020},
}

@ARTICLE{SAEJ3216,
author={SAE}, 
journal={{SAE} {J3216} },
title={Taxonomy and Definitions for Terms Related to Cooperative Driving Automation for On-Road Motor Vehicles},
year={2020},
}

@inproceedings{dresner2004multiagent,
  title={Multiagent traffic management: A reservation-based intersection control mechanism},
  author={Dresner, Kurt and Stone, Peter},
  booktitle={Autonomous Agents and Multiagent Systems, International Joint Conference on},
  volume={3},
  pages={530--537},
  year={2004},
  organization={Citeseer}
}

@article{malikopoulos2018decentralized,
  title={A decentralized energy-optimal control framework for connected automated vehicles at signal-free intersections},
  author={Malikopoulos, Andreas A and Cassandras, Christos G and Zhang, Yue J},
  journal={Automatica},
  volume={93},
  pages={244--256},
  year={2018},
  publisher={Elsevier}
}

@article{zhang2019decentralized,
  title={Decentralized optimal control of connected automated vehicles at signal-free intersections including comfort-constrained turns and safety guarantees},
  author={Zhang, Yue and Cassandras, Christos G},
  journal={Automatica},
  volume={109},
  pages={108563},
  year={2019},
  publisher={Elsevier}
}

@inproceedings{zhang2018decentralized,
  title={A decentralized optimal control framework for connected automated vehicles at urban intersections with dynamic resequencing},
  author={Zhang, Yue and Cassandras, Christos G},
  booktitle={2018 IEEE Conference on Decision and Control (CDC)},
  pages={217--222},
  year={2018},
  organization={IEEE}
}

@article{wang2021digital,
  title={Digital twin-assisted cooperative driving at non-signalized intersections},
  author={Wang, Ziran and Han, Kyungtae and Tiwari, Prashant},
  journal={IEEE Transactions on Intelligent Vehicles},
  volume={7},
  number={2},
  pages={198--209},
  year={2021},
  publisher={IEEE}
}

@INPROCEEDINGS{SUMO2018,
  author={Lopez, Pablo Alvarez and Behrisch, Michael and Bieker-Walz, Laura and Erdmann, Jakob and Flötteröd, Yun-Pang and Hilbrich, Robert and Lücken, Leonhard and Rummel, Johannes and Wagner, Peter and Wie{\ss}ner, Evamarie},
  booktitle={2018 21st International Conference on Intelligent Transportation Systems (ITSC)}, 
  title={Microscopic Traffic Simulation using SUMO}, 
  year={2018},
  volume={},
  number={},
  pages={2575-2582},
  doi={10.1109/ITSC.2018.8569938}}

@INPROCEEDINGS{5982524,
  author={Katsaros, Konstantinos and Kernchen, Ralf and Dianati, Mehrdad and Rieck, David},
  booktitle={2011 7th International Wireless Communications and Mobile Computing Conference}, 
  title={Performance study of a Green Light Optimized Speed Advisory {(GLOSA)} application using an integrated cooperative {ITS} simulation platform}, 
  year={2011},
  volume={},
  number={},
  pages={918-923},
  doi={10.1109/IWCMC.2011.5982524}}

@ARTICLE{6058657,
  author={Ferreira, Michel and d'Orey, Pedro M.},
  journal={IEEE Transactions on Intelligent Transportation Systems}, 
  title={On the Impact of Virtual Traffic Lights on Carbon Emissions Mitigation}, 
  year={2012},
  volume={13},
  number={1},
  pages={284-295},
  doi={10.1109/TITS.2011.2169791}}

@article{JAME2022167,
title = {Reducing pollutant emissions through Virtual Traffic Lights},
journal = {Computer Communications},
volume = {188},
pages = {167-177},
year = {2022},
issn = {0140-3664},
doi = {https://doi.org/10.1016/j.comcom.2022.03.018},
url = {https://www.sciencedirect.com/science/article/pii/S0140366422000937},
author = {Ahmadreza Jame and Marco Rapelli and Claudio Casetti},
}

@article{BAZZI201642,
title = {A distributed virtual traffic light algorithm exploiting short range {V2V} communications},
journal = {Ad Hoc Networks},
volume = {49},
pages = {42-57},
year = {2016},
issn = {1570-8705},
doi = {https://doi.org/10.1016/j.adhoc.2016.06.006},
url = {https://www.sciencedirect.com/science/article/pii/S1570870516301548},
author = {Alessandro Bazzi and Alberto Zanella and Barbara M. Masini},
}
